# Implementation Of Fuzzy-C4.5 Classification As a Decision Support For Students Choice Of Major Specialization

[1] Harsiti, [2] Tb. Ai Munandar, [3] Haris Triono Sigit

[1] Information System Dept., [2,3] Informatics Engineering Dept. Universitas Serang Raya
Banten-INDONESIA

## Abstract

*Determination of major specialization is important to lead the student to focus on areas of study that are of interest as well as in accordance with its academic credentials. Currently, the determination of major specialization is done by asking directly to students, regardless of academic outcomes that have been achieved in the previo us semester. This study discusses the development of a hybrid model from fuzzy-Mamdani and C4.5 algorithm to analyze the determination of major specialization in informatics engineering courses of Universities Raya Serang, where C4.5 algorithm is used as a shaper rule (rule) which is used in the inference stage. Establishment of rules (decision tree) performed using Weka applications, while for the determination of the decision support analysis specialization majors using Mamdani fuzzy concept, the application is done using the help of MATLAB. The results showed that 17 of the 126 students who either choose according to variable concentrations used in this study.*

## 1. Introduction

Elections for specialization majors among college students is important and not easy because it is not just based on sheer desire factor, but supported by student achieve academic proficiency. Selection of appropriate specialization course can serve as a strong foundation for students to define research topics at the end of the semester [1].

Recognizing the ability of students in accordance with the field and expertise is an important factor in determining the worth of a student choosing a particular concentration. This ability can be recognized by looking at the academic value of certain subjects at the core of major concentration [2]. Many cases where students feel in the wrong concentration so chose to move the concentration of middle class, academic grades achieved due to inadequate concentration election is not uncommon to just follow what his friend said, without promoting academic abilities possessed.

Furthermore, when students are exposed to concentrations of course, was shocked and unprepared and find it difficult to define the research topic as a final project. This is what lies behind the selection of this research topic.

Various methods and approaches can be used to make determinations of specialization majors, one of which is a fuzzy inference mamdani, where the inference stage, requires a few rules that have been established to determine the final outcome in stage defuzzyfication. Establishment of rules is done by using the C4.5 algorithm. This is done because there is currently no criteria are owned by the Department of Informatics, University of Raya Serang to establish rules determining specialization majors. By using eight core subjects variable and value-adjusted performance index of student data that has been concentrated, analyzed the data using Weka application to obtain the rules that will be used. This study discusses the implementation of fuzzy inference mamdani and C4.5 algorithms for determining specialization for students majoring in Information Engineering University Raya Serang.

## 2. Fuzzy-Mamdani

Fuzzy logic inference systems have some models that can be used to browse or reasoning with principles such as human reasoning process approach. The model are, Tsukamoto, Mamdani and TSK (Takagi Sugeno Kang). In practice, fuzzy Mamdani more widely used as tools reasoning process of a system is built. This is because each step of the work going on, closer to the human way of thinking and reasoning in doing the work according to the rules of linguistics.

Mamdani fuzzy output obtained through four major phases, namely, fuzzyfikasi stage, the formation of fuzzy knowledge bases, applications and implications defuzzyfication function [3], [4], [5], [6]. Various studies and applications developed using Mamdani, including the determination of groundwater quality analysis [7], the prediction results of the general election (election) [8], the determination of channel estimates for the case of a wireless communication network signal is reduced [9], modeling the conditions for laser machine determine the level of surface roughness on the lasering process [10], modeling the auto zoom on digital cameras [11], the determination date of the evaluation process quality grading of fruit to be exported [12], as well as its use for the concept konwlegde management system [13].

Fuzzy logic has a membership function which indicates the degree of a set of the universes of discourse. Membership functions allows us to represent fuzzy sets into a graphical form. There are many membership function graphs can be used to determine the membership functions in a fuzzy set, namely, the function of triangular, trapezoidal functions and Gauss functions.

## 3. C4.5 Algorithm

C4.5 algorithm is an extension of ID3 algorithm so that it works the same as the basic principle of the algorithm. C4.5 algorithm selected as forming rules for rule evaluation in fuzzy method, this algorithm is able to handle the attributes that have continuous data and discrete, capable of handling the data that have missing value, the results of the decision tree can be pruned once formed, thus enabling retrieval in accordance with the rules attributes are used, in addition to the algorithm C4.5 uses gain ratio to perform the selection of attributes that will serve as a branch of the decision tree to be formed.

C4.5 algorithm is an algorithm that can be used for the establishment of a decision tree according to attributes that are divided into smaller subsets, where the process of forming a decision tree or rule Decission influenced by information gainnya value [14]. C4.5 algorithm has the advantage that it is able to extract hidden information from large data sets and produce the expected classification rules with very good accuracy [14], [15], [16], [17].

C4.5 has been widely used for a variety of needs for both academic and research interests are implemented in a real case. Research on classification of speech using a particular audio frame [18], the determination of one's talents prediction [19], classification of micro-blogging kind of questions to find a correlation between teacher questions in class on a given course material [20] and application monitoring for device connectivity network [21] is a C4.5 algorithm implementation. In general, the decision tree algorithm C4.5 form by following the following order:

1. Select attributes as the root
2. Create a branch for each value
3. For cases in the branch
4. Repeat the process for each branch until all cases the branches have the same class.

Highest gain is used for the selection of attributes as the root, with the following equation:

$$Gain(S, A) = Entropy(S) - \sum_{i=1}^{n} \frac{|S_i|}{|S|} * Entropy(S) \quad (1)$$

Description:
S: the set of cases
A: Attributes
N: number of attribute partition A
Si: number of cases in the i-th partition
S: the number of cases in S

While the value of Entropy, sought by the following equation:

$$Entropy(S) = \sum_{i=1}^{n} -p_i * log_2 * p_i \quad (2)$$

Description:
S: the set of cases
n: number of partitions S
pi: the proportion of the S

## 4. Fuzzy-C4.5 Model

Fuzzy-C4.5 models are referred to in this research is, how the results of the analysis are formed using a decision tree C4.5, used as a rule to be evaluated in rule evaluation stage after stage fuzzyfikasi. Rules that form is the result of data extraction current students have chosen majors, so the determination of rules put forward the real data is available. The following diagram Fuzzy-C4.5 models:

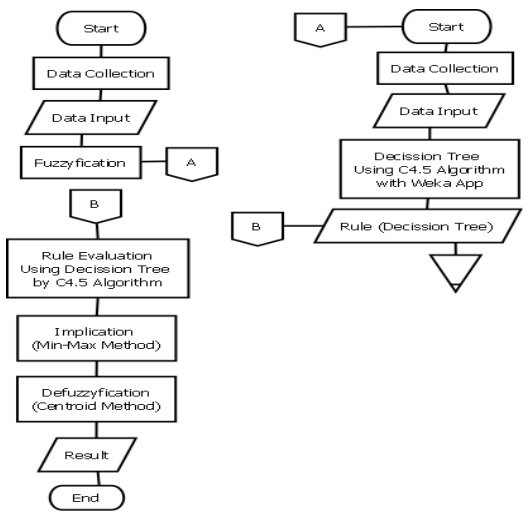

Figur 1. Fuzzy-C4.5 Model

## 5. Research Methodology

This study begins with the value of student data collection, and produces a number of data and then do the selection and data cleaning to produce as many as 126 data to be used as research material. A total of 13 variable data used in this study, including 8 course grades, GPA 4 value data, and a classification variable specialization. The data was then analyzed to determine a match, so it can be used for Fuzzy-C4.5. The next step is to form a decision tree (rules) using Weka application, produced as many as 16 rules. Once the rules are known, then analyzed using a Mamdani fuzzy MATLAB applications, and include rules that have been formed into a Mamdani fuzzy inference engine in MATLAB.

## 6. Result and Discussion

Prior to analysis using a fuzzy-Mamdani, the first step is to extract the data to form a decision tree using Weka applications. The data have been extracted generating 16 rules of decision tree is formed (see figure 2)

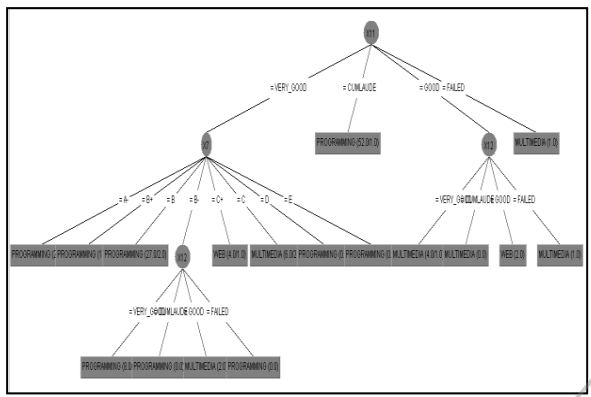

Figure 2. Decission Tree Using C4.5 Algorithm

```
X11 = VERY_GOOD
|   X7 = A-: PROGRAMMING (2.0/1.0)
|   X7 = B+: PROGRAMMING (17.0/5.0)
|   X7 = B: PROGRAMMING (27.0/2.0)
|   X7 = B-
|   |   X12 = VERY_GOOD: PROGRAMMING (8.0/2.0)
|   |   X12 = CUMLAUDE: PROGRAMMING (0.0)
|   |   X12 = GOOD: MULTIMEDIA (2.0)
|   |   X12 = FAILED: PROGRAMMING (0.0)
|   X7 = C+: WEB (4.0/1.0)
|   X7 = C: MULTIMEDIA (6.0/2.0)
|   X7 = D: PROGRAMMING (0.0)
|   X7 = E: PROGRAMMING (0.0)
X11 = CUMLAUDE: PROGRAMMING (52.0/1.0)
X11 = GOOD
|   X12 = VERY_GOOD: MULTIMEDIA (4.0/1.0)
|   X12 = CUMLAUDE: MULTIMEDIA (0.0)
|   X12 = GOOD: WEB (2.0)
|   X12 = FAILED: MULTIMEDIA (1.0)
X11 = FAILED: MULTIMEDIA (1.0)
```

Figure 3. Decission Tree Using C4.5 Algorithm

The results of the decision tree that is formed is then used as a rule in rule evaluation stage after stage fuzzyfikasi. The following is an inference engine that forms on MATLAB applications (see Figure 4).

Inference engine has 12 input variables and 1 output variable.

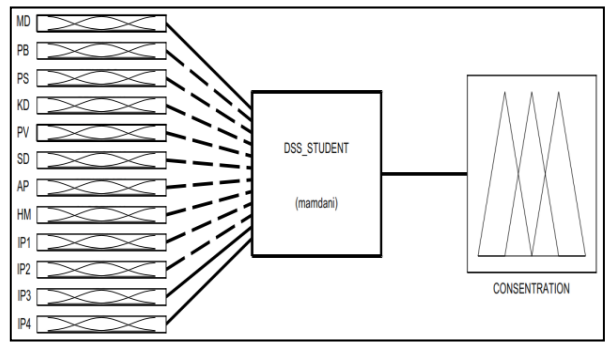

Figure 4. FIS of Student DSS Specialization Determination

Each input variable value data subjects, have a membership function as follows:

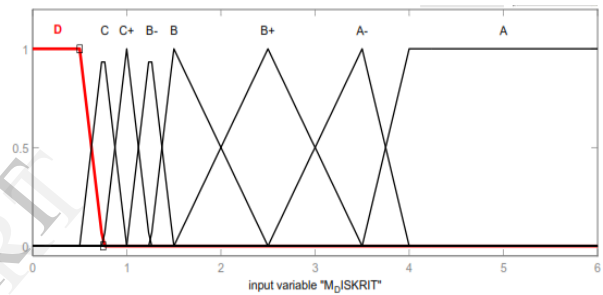

Figure 5. Membership function of course subject

As for the output variable has three sets namely concentration Multimedia, Web and Programming (see figure 6).

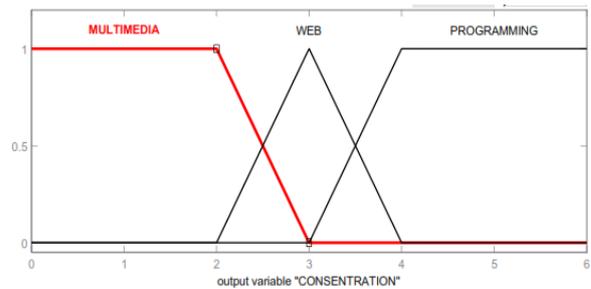

Figure 6. Membership function of Output

The following are the rules used for the evaluation phase of the fuzzy inference rules on engine selection decision support mamdani concentration majors (see figure 7).

Once the inference engine is formed according to the case at hand, further testing against the data to 126 students who have specialization majors. Defuzzyfication method used in this case is the centroid method. From the test results obtained that the suitability analysis using fuzzy inference by 86.51% with a mismatch rate of 13.49%. Or as many as 17 students who choose majors based on academic grades should not be owned. Here's the data that indicated students choose majors that are not appropriate concentration (see table 1).

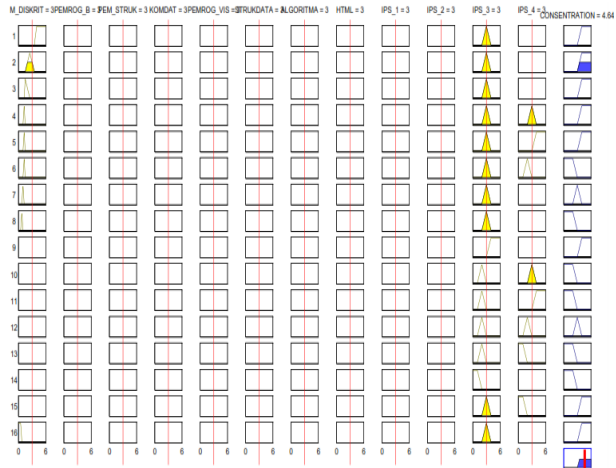

Figure 7. Rule on FIS DSS

Here's the data that indicated students choose majors that are not appropriate concentration (see table 1).

Table 1. Analyzed Student Data

| NO | STUDENT_NAME | CURENT_MAJOR | RECOMMENDATION |
|---|---|---|---|
| 1 | SRI RAHAYU M. | MULTIMEDIA | PROGRAMMING |
| 2 | RICKY SUPRIYADI | MULTIMEDIA | PROGRAMMING |
| 3 | IFAN MUKHLISIN | MULTIMEDIA | PROGRAMMING |
| 4 | KURNIAWAN | MULTIMEDIA | PROGRAMMING |
| 5 | DARA ASTUTI | MULTIMEDIA | PROGRAMMING |
| 6 | SYAHDUDIANI | MULTIMEDIA | PROGRAMMING |
| 7 | IRVAN RAMDHANI | MULTIMEDIA | PROGRAMMING |
| 8 | ADIYANTO | MULTIMEDIA | WEB |
| 9 | HUMAIDI | MULTIMEDIA | PROGRAMMING |
| 10 | KIKI RIZKI RAMADAN | MULTIMEDIA | PROGRAMMING |
| 11 | AHMAD MUJIBI | MULTIMEDIA | PROGRAMMING |
| 12 | ANDRI FITRIANSYAH | MULTIMEDIA | PROGRAMMING |
| 13 | M. HIKMATULLAH | WEB | PROGRAMMING |
| 14 | UNUNG EDIYANA | WEB | PROGRAMMING |
| 15 | DONA NINGSIH | WEB | PROGRAMMING |
| 16 | DEFFA JENIVER | WEB | PROGRAMMING |
| 17 | AHMAD KOHARSYAH | WEB | PROGRAMMING |

## 7. Conclusion

Based on the analysis make use of Fuzzy-C4.5 can be obtained by 17 students who indicated specialization majors chose not appropriate based on its academic value. The test results indicate that the inference engine matches the level of accuracy that is made, at 86.51%. The results could certainly be one of the decision support for the determination of concentration for students majoring in accordance with its academic value.